\documentclass[preprint]{aastex}

\newcommand{\Teff}{$T_{\rm eff}$}
\newcommand{\vt}{$v_{\rm t}$}
\newcommand{\logg}{log~$g$}
\newcommand{\lgf}{log $gf$}

\newcommand{\aprox}{${\sim}$}
\newcommand{\kms}{km s$^{-1}$}
\newcommand{\nd}{\nodata}
\newcommand{\stdv}{$1{\sigma}$}

\received{}
\accepted{}

\shortauthors{Cavallo, Suntzeff, \& Pilachowski }
\shorttitle{Al in Cluster Giants}

\begin{document}

\title{Hydra Observations of Aluminum Abundances in the Red Giants of the Globular Clusters
 M80 and NGC~6752}

\author{Robert M. Cavallo\altaffilmark{1}}
\affil{Lawrence Livermore National Laboratory, 7000 East Ave, L-97, Livermore, CA 94450}
\email{rcavallo@llnl.gov}

\author{Nicholas B. Suntzeff}
\affil{Cerro Tololo Inter-American Observatory}
\affil{National Optical Astronomy Observatory}
\affil{Casilla 603, La Serena, Chile}

\and

\author{Catherine A. Pilachowski}
\affil{Astronomy Department, Indiana University, Swain West 319, 727 E. 3rd St, Bloomington, IN 47405}

\altaffiltext{1}{Visiting Astronomer, Cerro Tololo Inter-American
Observatory, National Optical Astronomy Observatory, which is operated
by the Association of Universities for Research in Astronomy,
Inc. (AURA) under cooperative agreement with the National Science
Foundation.}

\begin{abstract}
Aluminum and other metal abundances were determined in 21 red giants
 in the globular clusters NGC~6752 and M80 as part of a larger study
 to determine whether the aluminum distribution on the red giant branch
 is related to the second parameter effect that causes clusters of similar
 metallicity to display different horizontal branch morphologies.
The observations were obtained of the Al~I lines near 6700 {\AA} with the CTIO
 Blanco 4-m telescope and Hydra multi-object spectrograph.
The spectra have a resolving power of 18000 or 9400, with typical S/N ratios
 of 100-200.
Mean [Fe/H] values obtained from the spectra are --1.58 for NGC~6752
 and --1.73 for M80; this represents the first spectroscopic iron abundance
 determination for M80.
Both NGC~6752 and M80 display a spread in aluminum abundance, with mean
 [Al/Fe] ratios of +0.51 and +0.37, respectively.  No trend in the
 variation of the mean Al abundance with position on the giant branch is
 discernible in either cluster with our small sample.
\end{abstract}


\section{INTRODUCTION}
\label{sec:intro}

Over 20 years have passed since \citet{NCFD81} first showed aluminum
 abundance variations on the red-giant branch (RGB) of the
 globular cluster NGC~6752, and yet the nature of these inhomogeneities 
 remains a mystery.
Observations since then continue to show aluminum (and sodium and magnesium) variations
 in other clusters \citep[see, e.g.][]{CD81,WLO87,SHFS87,DSS92,
 ND95,PSKL96,ZWB96,S96a,Kraft97,SKS97,K98,Ivans1999,CN2000,Ivans2001,RC2001,GratAl2001,
 GBNF2002}.
Although variations in carbon and nitrogen were previously known, these
 could be described through a simple mixing mechanism proposed by
 \citet{SM79}, where rotationally induced meridional circulation
 currents could carry nuclearly processed materials such as C, N, and
 O from around the hydrogen-burning shell (H shell) of a red giant across the
 radiative zone to the outer convective envelope.
Heavier elements such as Mg, Na, and Al weren't thought to be
 processed around the H shell, and any variations in them were taken as
 evidence that some kind of primordial pollution affected the
 surface abundances \citep[see, e.g.,][]{CD81}.
Continued work on key nuclear reaction rates \citep{Champagne98,DD90,Iliadis90,
 CBS93,Blackmon95,Iliadis96}, however, suggested
 that these elements could be processed around the H shell under
 the same conditions that the CN and ON nuclear cycles operated.
Using these results, as well as the widely accepted rates of \citet{CF88},
 separate groups showed that it is possible to account qualitatively
 for the observed variations that showed Na, Al, and N anticorrelated with
 C, O, and, in some cases, Mg \citep{LHS93, CSB96, DW96, CSB98, DDNW98}.
The challenge has always been describing the results {\em quantitatively};
 in particular, producing [Al/Fe] as high as 1.5 dex without
 overproducing [Na/Fe], and depleting $^{24}$Mg to the observed levels
 in M13 \citep{S96b}, all while remaining within the acceptable proton-capture
 rates \citep{LHZ97,NACRE,Powell99,UNC}.
Despite any latitude afforded by the reaction rate uncertainties 
 \citep[see, e.g.,][]{CSB98}, the mixing theory still relies on non-solar
 abundance ratios in the star prior to mixing \citep{DDNW98,CN2000}, thus,
 at least partly relying on primordial influences.
Theories other than meridional circulation have been put forth \citep{LHZ97,FAK99,
 AFK2001,DW2001}, but a detailed discussion of each is beyond the intent of
 this paper.

Regardless of the physics behind any mixing mechanism, the question still
 remains: are the Al (and Na, Mg) variations caused by mixing, primordial
 sources, or a combination of both?
We might be able to answer this question by looking at another
 long outstanding problem in globular cluster astronomy, namely, the
 second-parameter effect.
First pointed out by \citet{SW67} and \citet{vdB67}, the second-parameter effect
 refers to the phenomenon where the horizontal-branches (HB) of two clusters with similar
 metallicity (the first parameter) have markedly different color distributions.
%Well-known pairs of second-parameter clusters include M13/M3 and NGC~288/NGC~362,
% for example.
Possible second parameters that have been investigated include 
 age, initial helium abundance, CNO abundance, and mass loss
 \citep[see, e.g.,][]{Faulkner66,Renzini77,Chaboyer98}, among others,
 but none is applicable to all clusters.
One recent suggestion hypothesizes that if deep mixing (i.e., mixing that
 penetrates the H shell) occurs, then helium will be brought to the
 surface affecting the HB morphology by making mixed stars both bluer
 and brighter (Sweigart 1997a,b). 

Unfortunately, helium cannot be measured in RGB stars because of their low
 surface temperatures, and helium settling on the HB precludes an accurate
 measurement in the hotter stars.
However, models by \citet{CSB98}, which use standard (albeit, uncertain) 
 reaction rates, show that aluminum can be produced only in the H shell of
 giants within the last magnitude of the RGB, implying
 that an increase in helium in the envelope must also be accompanied by an
 increase in aluminum.
From this we can postulate that, if [Al/Fe] variations are produced internally
 and not primordially, Al could be a good surrogate to measure He mixing
 on the bright RGB, and if He mixing is indeed the second parameter, a relationship should
 exist between the ratio of Al-strong to Al-normal stars (the Al ratio) and
 the ratio of blue to red HB stars (the HB ratio).
It is the {\em distribution} of abundances as a function of magnitude that is
 critical in determining the whether this correlation exists and what its cause
 might be.

This current work attempts to provide some fresh data on clusters that have been
 historically under-studied.
Both M80 (NGC~6093) and NGC~6752 possess blue HBs relative to other clusters
 at similar metallicity.
For example, \citet{Ferr98} compared Hubble $U,V$ M80 photometry directly
 with M13 and M3, with the result that M80's HB is very similar to M13, while
 M3 lacked the extended blue tail of the other two clusters.
Meanwhile, \citet{GCLSA1999} presented extensive Str\"{o}mgren photometry of NGC~6752,
 M13, and M3, showing extensive blue tails in the former two compared
 with the latter.
Neither M80 nor NGC~6752 had been studied extensively for abundances until
 \citet{GratAl2001} and \citet{GBNF2002}
 determined [Al/Fe] for 39 stars in NGC~6752 in total, extending the data of
 \citet{ND95}.
These were significant results because they probed less evolved stars near
 the main-sequence turnoff, on the subgiant branch, and at the base of the RGB,
 which are below the point that mixing theories predict that aluminum
 can be produced.
Using a non-LTE analysis, \citet{GratAl2001} observed dwarfs with [Al/Fe] as low as $-0.76$ 
 dex from the Al~I resonance lines, and subgiants with [Al/Fe] as high as $+0.86$
 dex from the doublet at ${\lambda}{\lambda}$8773/74~{\AA}.
In fact, the resonance line analysis for the dwarfs yielded [Al/Fe]~$=~-0.18~{\pm}~0.15$~(s.e.m.)~dex,
 while the subgiants gave [Al/Fe]~$=~+0.29~{\pm}~0.11$~dex (s.e.m.).
The results for the dwarfs are uncertain due to the difficulty of analyzing resonance
 lines with non-LTE corrections of as much as $+0.6$~dex; yet, the results are
 still surprisingly low.
What causes the drastic change from the main sequence to the subgiant branch: atmospheric
 effects, the different choice of lines, or an actual physical phenomenon?
The \citet{GBNF2002} results are more in line with the results of \citet{ND95} and
 with other clusters.
We discuss implications of the aluminum data in NGC~6752 in more detail in
section~\ref{sec:final_look}.

M80, on the other hand, has no published abundances and is in need of further
 investigation, especially given its similarities to M13 in metallicity and HB morphology.

The rest of the paper is outlined according to the following:
We begin with a description of the observations in Section~\ref{sec:obs}, followed
 by our data reduction techniques in Section~\ref{sec:ccd}.
We then discuss membership criteria in Section~\ref{sec:rv}.
After culling the data, we show the results of our abundance analysis in 
Section~\ref{sec:analysis} and give our final conclusions
 in Section~\ref{sec:conclude}.

\section{OBSERVATIONS}
\label{sec:obs}

Between 1999 and 2001 we used the CTIO Blanco 4m telescope with the Hydra multi-object
 spectrograph in the echelle mode to observe 105 stars near NGC~6752 and 47 stars near M80.
In the end, only a subset of these spectra were of sufficient quality to determine
 both cluster membership and abundance information.
We describe the observations in the following two subsections, and discuss the radial 
 velocity membership criteria in Section~\ref{sec:rv}.

\subsection{Observing Parameters and Instrument Configurations}
\label{sec:ob_parms}

Our observations were taken during two different epochs, between
 which the instrument underwent substantial changes.
The first run was during the commissioning phase of Hydra in
 1999, and only concentrated on NGC 6752 giants, while the second
 run occurred in 2001, and involved both clusters.
The different instrument parameters for each epoch are listed in 
 Table~\ref{tab:t01}.
The net effect of the change between the two runs is the higher
 resolution offered in 2001 by placing slit plates in front of
 the large fibers and using the longer focal-length camera.
The resolution was measured using the FWHM of narrow emission lines
 in the comparison lamps.
Both observing runs utilized the echelle grating.

\placetable{tab:t01}

\subsection{Targets}
\label{sec:targs}

Comprehensive membership surveys were available for neither M80 nor NGC 6752
when stars were selected for observation with Hydra,
limiting our ability to ensure that fibers were assigned
  to actual cluster members and not to field stars.
Both clusters are located at relatively low galactic latitude (+19 degrees for M80 and 
 --25 degrees for NGC 6752), and field star contamination is likely without proper
 motion or radial velocity information.
For NGC 6752, we used the $B,V$ photographic photometry of \citet{ngc6752} to 
 select likely cluster members.
This cluster has a large tidal radius of 55.{\arcmin}34 \citep{Harris96},
 compared with the 40{\arcmin} field of view for Hydra, ensuring that most stars
 in the aperture were likely to be members.

Since no wide-field photometry exists for M80, we obtained $B,V$ CCD
 photometry of nine 13{\arcmin}${\times}$13{\arcmin} fields centered on M80 
 with the CTIO 0.9m telescope prior to our Hydra run.
The weather was not photometric, but
 we were able to obtain reasonable estimates of colors (${\pm}$ 0{\fm}05), and
 accurate astrometry after tying the system to the USNO-A2.0 catalog \citep{USNO}.

As a best guess for determining actual members of both clusters prior to our runs,
 we combined the existing photometry with the available proper motion data.
For both clusters, we received accurate astrometric data from D. Dinescu (1998, 2001,
 private communication), which also included proper motion information.
The average proper motions for 998 NGC 6752 stars in the Dinescu sample were
 ${\mu}_{\alpha}$~=~$-0.5~{\pm}~23.0$~({\stdv})~mas~yr$^{-1}$, and 
 ${\mu}_{\delta}$~=~$+3.0~{\pm}~31.0$~({\stdv})~mas~yr$^{-1}$, 
 around the cluster center.
We selected for observation only stars with $-14.2~{\le}~{\mu}_{\alpha}~{\le}~+13.7$~mas~yr$^{-1}$,
 and $-9.8~{\le}~{\mu}_{\delta}~{\le}~+9.9$~mas~yr$^{-1}$ for observing.
Dinescu's data for M80 were for stars in an annulus far outside the small cluster, and were
 less useful for selecting probable members.

The final lists of stars that we ultimately determined to be cluster members and that
 had spectra with sufficient S/N to analyze reliably are given in 
 Tables~\ref{tab:t02} and \ref{tab:t03}.
The star identification in the first column in Table~\ref{tab:t02}, as well as the photometry for
 the NGC~6752 giants are from \citet{ngc6752}, while the alternate name in column 2 is from
 Dinescu's work.
The colors for NGC~6752 are corrected assuming E($B~-~V$) = 0.04, as listed by \citet{ngc6752}.
The exposure times and S/N near 6700~{\AA} are listed for
 each observing run separately.
For M80, the star number and photometry are our own\footnote{The astrometry is available 
 from the first author.}, with E($B~-~V)~=~0.17$ magnitudes assumed \citep{KG76,RHS88}.
Despite the rather large list of stars that we observed, only 21 made it past the
 cutting room floor.
The main obstacle to obtaining more spectra was the need for both high S/N and
 high resolution, which was aggravated by varying observing conditions and
 intermittent mechanical difficulties during the commissioning of Hydra.

\placetable{tab:t02}
\placetable{tab:t03}

\section{CCD PROCESSING AND SPECTRA EXTRACTION}
\label{sec:ccd}

The data from both epochs were processed in similar fashion using
 the reduction routines in IRAF \citep{IRAF}, beginning with
the usual overscan and bias corrections.
We made our flat-field image by combining several daytime sky flats that were
 exposed by using diffusing filter in place of the echelle filter,
 removing the spectral shape in the x and y directions by curve-fitting,
 and boxcar filtering the remainder, which left us with a
 smooth milky flat that contained only pixel-to-pixel variations.
The data images were then divided by this flat.
The object spectra were extracted using the IRAF task APALL with
 variance weighting, background subtraction, and cosmic-ray cleaning
 parameters turned on.
The fiber-to-fiber response functions were calculated using the
 MSRESP1D task, where the individual fiber responses in an averaged
 quartz flat were determined with reference to an averaged twilight
 sky flat.
To remove fiber-to-fiber variations as well as most of the
 instrument profile, these responses were then divided through the extracted 
 spectra to create normalized spectra.

To determine the dispersion solutions for both epochs,
 we used etalon exposures bracketed around our object
 exposures throughout the night.
These nighttime etalon spectra were calibrated with ``master'' etalon spectra
 that were acquired early during the first night of each run and 
 were themselves calibrated from a ThAr lamp for the first epoch, and
 a HeNeAr lamp for the second epoch.
The dispersion solution for a gas-filled lamp was applied to the extracted master
 etalon spectra, which were then averaged together to form one spectrum.
The individual lines in the averaged master etalon spectrum were then measured 
 and used to calibrate the individual spectra from the master etalon image
 so that every fiber used the same set of lines in its dispersion solution.
These master etalon spectra were then used to calibrate the nighttime etalon
 spectra.
We checked the quality of our etalon solutions by comparing ones
 taken on two different nights and found that both fiber-to-fiber and
 night-to-night dispersion solutions remained consistent to less than
 0.002~{\AA}.

We prefer using the etalon because of its superior precision.
For example, during the second epoch the HeNeAr exposure only contained 
 17 usable lines over {\aprox}300 {\AA}, with most lines on the red side 
 of the spectrum, while the etalon
 contained 95 strong unblended lines uniformly spaced over the spectrum.
The typical rms for the HeNeAr lamp was around 0.03 {\AA} while the etalons
 were better by an order of magnitude.

Sky spectra were obtained by pointing {\aprox}20 of the available fibers at empty fields
 during the program star exposures.
After extraction and calibration, the skies were averaged together
 to build a clean single background spectrum, which was subtracted from the
 object spectra.
Finally, the program star spectra were flattened with a high order spline
 that was fitted through the continuum.
No telluric line corrections were made since there were no telluric lines
 in our spectra during the second epoch and the ones appearing in the earlier
 data are outside our window of interest.

\section{RADIAL VELOCITIES}
\label{sec:rv}

After extracting and calibrating the spectra, we measured radial velocities in
 order to determine cluster membership.
While the etalon lamps gave very high precision wavelength calibrations,
 they are uncertain in absolute wavelength.  
Raw radial velocities were determined for each star using strong lines in our
highest S/N ratio spectra.
We calibrated the velocity scale for each cluster observation by setting the mode
of the velocity distribution equal to the published radial velocity, using
 $+8.2~{\pm}~1.5$~({\stdv})~{\kms} for M80 and
 $-27.9~{\pm}~0.8$~({\stdv})~{\kms} for NGC 6752 \citep{Harris96}.
We accepted for further study all stars that that had radial velocities
within 22 and 15~{\kms} of the cluster means for M80 and NGC 6752, respectively, 
and which have sufficient S/N ratio for reliable abundance determinations (typically
above 100 for epoch 1 and 60 for epoch 2).

\section{ABUNDANCE ANALYSIS}
\label{sec:analysis}

\subsection{Equivalent Widths}
\label{sec:ew}

We determined a set of lines to measure by examining the spectra with the
 highest S/N from each set of cluster giants.
The lines were identified and compared with the solar spectrum of Wallace,
 Hinkle, \& Livingston (1993), and with Table~II of \citet{Thev90},
 from which the oscillator strengths were chosen, with the following exceptions:
 the wavelengths and excitation potentials of the Fe I lines come from \citet{Nave94},
 and we use the atomic data for the Eu~II line at {$\lambda$}6645~{\AA} that is
 consistent with hyperfine splitting, as described in sec~\ref{sec:results}.
This produced an Eu oscillator strength that is +0.22~dex stronger than the value listed by
 \citet{S96a} and is $+0.51$ dex stronger than the \cite{Thev90} value.
For the other lines, we used only those that \citet{Thev90} cites as having an 
 uncertainty in {\lgf} of less than 0.05 dex, and stayed away from strong blends where 
 the separate cores of each line could not be individually detected in our spectra.
We measured equivalent widths using Gaussian fits with the IRAF routine SPLOT
 and made use of the SPLOT deblending algorithm for lines that overlapped,
 but appeared visually distinct.
The line list and equivalent widths are given in Table~\ref{tab:t04}
 for NGC~6752, and Table~\ref{tab:t05} for M80.

\placetable{tab:t04}
\placetable{tab:t05}

\begin{figure}[htb]
\epsscale{0.5}
\plotone{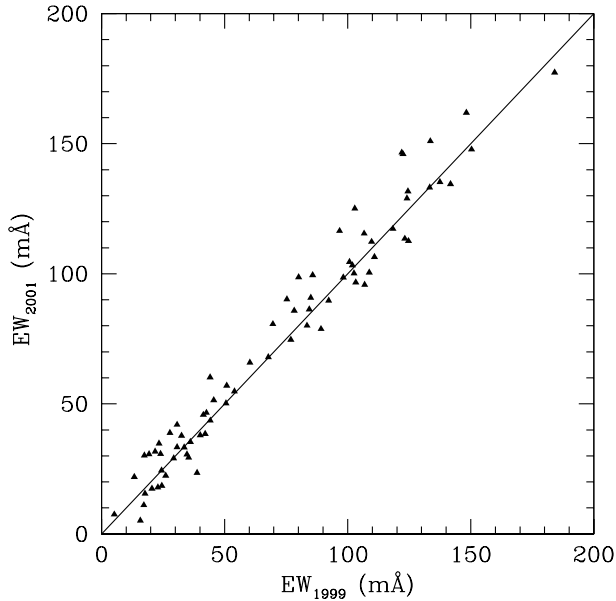}
\caption{Comparison of equivalent widths from a sample of spectra observed
         in both 1999 and 2001.  The diagonal line represents perfect 
	 agreement.}
\label{fig:f01}
\end{figure}

To check for consistency from epoch to epoch, we measured the
 equivalent widths of 12 lines in the spectra of seven NGC 6752 giants that were
 observed during both runs, and show the results in Figure~\ref{fig:f01}, where the
 solid line represents perfect agreement.
The average difference between the two datasets is $2.61~{\pm}~8.89$ m{\AA}
 (1~${\sigma}$), in the sense of the 2001 data minus the 1999 data.
Considering the different instrument capabilities between the two epochs
 and the different S/N quality of the datasets, we consider this to be good agreement.

\subsection{Model Parameters: {\Teff}, Gravity and Microturbulence}
\label{sec:mods}

We explored two methods of determining model atmosphere parameters: spectroscopically
 from Fe~I lines, and photometrically from $B,V$ colors.
Each method proved equally challenging, yet both provided consistent results.

To derive atmospheric models from Fe~I lines, we employed the usual approach of
 removing dependencies of the derived abundances on both the excitation
 potentials and the reduced widths\footnote{Reduced width~=~equivalent width divided
 by wavelength.} to derive {\Teff} and {\vt}, respectively.
We used MOOG \citep[, version 2000]{MOOG}, the LTE analysis routine, in conjunction
 with MARCS \citep{MARCS}, the plane-parallel stellar atmosphere code, to determine
 all the abundances in this paper.
The gravities were determined by interpolating between effective temperatures
 in a 12 Gyr isochrone with [Fe/H]$~=~-1.54$ and $[{\alpha}/{\rm Fe}]~=~+0.3$,
 provided to us by R. Bell [1999, private communication; see Houdashelt, Bell, \&
 Sweigart (2000)].
The difficulty with this approach is two-fold:  first there are only a handful
 of Fe~I lines available in the spectra, and usually only enough in the
 brightest (and coolest) stars to determine parameters.
Second, recent work has revealed possible problems with determining [Fe/H]
 using Fe~I lines due to NLTE effects \citep{TI99}, which
 are most exaggerated at lower metallicities and for low excitation-potential lines.
The \citet{TI99} results, however, are from main-sequence turn-off stars and
 it appears from their work and from \citet{KI2003} that the ``overionization''
 problem that might cause difficulties in less evolved stars probably isn't
 a concern here.

We also derived {\Teff} photometrically with the color-corrected $B,V$ data from
 \citet{ngc6752} for NGC 6752, using the Bell isochrone discussed above, which
 provides transformed colors in $B$ and $V$.
From this {\Teff} we determined an appropriate {\logg} from the isochrone, and then
 found {\vt} via the empirical formula given by \citet{PSKL96}; i.e.,
 {\vt}~$=~-8.6{\times}10^{-4}T_{\rm eff}~+~5.6$~{\kms}.
This approach provides microturbulent velocities that are, on a star-by-star basis,
 systematically higher by $0.24~{\pm}~0.16${\stdv}~{\kms} than those determined via
 Fe~I lines.
As discussed below, this has little effect on the final results.
Comparing {\Teff} from the spectroscopy with {\Teff} from the photometry for
 11 NGC 6752 giants that had spectra with S/N$~>~65$ and at least 11 Fe~I lines
 over a range of excitation potential from 1.0 eV to 4.8 eV, we found an
 average difference of $-39~{\pm}~81$~({\stdv})~K, with the spectroscopic determination
 yielding the lower values on average.
Removing the most discrepant case, star 277, reduces {\Teff}$_{- \rm spectrum}$
$-$ {\Teff}$_{- \rm photometry}$ to $-17~{\pm}~48$~({\stdv})~K.
The agreement is quite good given the uncertainties in the photometry and 
 small number of Fe~I lines, indicating that either
 approach is sufficient for determining model atmosphere parameters.
A similar analysis for the M80 giants shows the spectroscopic temperatures to
 be $5~{\pm}~65$~({\stdv})~K lower on average than the photometric determinations.

The brightest giant studied in NGC 6752, star 3589 in the Buonanno photometry,
 also has a color that is both too red for our isochrone at its magnitude, and
 inconsistent with the next brightest giant, star 2403, which is only 0.1
 magnitudes below it on the RGB; however, {\Teff} determined from the spectrum
 of star 3589 is only 25~K cooler than star 2403, indicating that its color
 is either in error, and it most likely belongs on the bright RGB, or that it
 is a variable star.

The model atmosphere parameters for our giants are listed in Table~\ref{tab:t06} for both
 clusters, with the method chosen for deriving the parameters indicated in column 5.
Where possible, we used the photometrically derived parameters, but
 occasionally had to rely on the spectroscopic models when the photometry
 appeared questionable.
This preference allowed us to utilize spectra that had lower S/N
 ratios than would be required to derive {\Teff} and {\vt} spectroscopically.
Because of the lower resolution of the first epoch data, we still, nonetheless,
 needed high S/N (${\gtrsim}~110$) to derive abundances, 
 while a factor of two lower S/N was sufficient for the second-epoch data.

\placetable{tab:t06}

Finally, we examine the sensitivity of our abundances on our choice of model atmosphere
 parameters, as shown in Table~\ref{tab:t07}.
The top row shows a reasonable estimate for the uncertainties in {\Teff}, while giving
 even wider latitude in ${\Delta}${\logg} and ${\Delta}${\vt}.
The variations are determined using the model and data from star 2240 in NGC~6752, which
 has a {\Teff} that is near the median value for our sample.
Since the sensitivity to the parameters is not symmetric, we report in the table the worse
 result of either increasing or decreasing a given parameter.
Only the 2001 data are used in this test since the spectrum has both a full complement
 of lines and a high S/N.
The results are not surprising: {\Teff} plays the biggest role in the error budget,
 with the low E.~P. lines of Ti~I and Cr~I being the most sensitive, while the ionized 
 lines are little affected.
Except for Ti~II and Eu~II, an increase in {\Teff} leads to an increase in [abundance/H].
Uncertainties in {\logg} lead to only small errors for even the pressure-sensitive
 ions with some elements increasing with {\logg} (Al, Ca, Ti~I, and Cr) and others
 decreasing (Ti~II, Fe, Ni, La, and Eu).
Finally, uncertainties due to {\vt} are quite small, which is consistent with the many weak
 lines employed in our analysis.
Increasing {\vt} desaturates the line thereby increasing the derived abundance, which
 is what we find in our results, with [Al/Fe] having zero dependence within the
 {\vt} uncertainties.
Of course, the three parameters are not uncorrelated, so variations in one necessarily
 lead to variations in the others (as can be seen by the formula for {\vt}, which is
 a function of {\Teff}) and the sensitivities reported in Table~\ref{tab:t07} can be
 be modified to reflect this in the total uncertainty due to the atmospheric models.
Since the focus of this study is primarily on aluminum, we note that the
 model uncertainties introduce only ${\sim}$~0.1 dex uncertainty in the [Al/Fe] ratio.

\subsection{Results}
\label{sec:results}

In Table~\ref{tab:t08} we present the abundances of 11 stars in NGC~6752 and
 10 in M80.
These were accepted as cluster members after determining that their derived
 [Fe/H] ratios were consistent with the anticipated metallicities of each cluster,
 around $-1.5~{\rm to}~-1.6$~dex.
We estimate that the abundances are uncertain by about 0.15 dex after
 accounting for line-to-line scatter, uncertainties in oscillator strengths and
 atmospheric parameters, and instrumental errors.
The abundances for the NGC~6752 data are averaged together when available by giving
 the second-epoch results twice the weight of the first to account for the factor
 of two improvement in resolution.
With the exception of star 3589, the abundances for NGC~6752 from the the two epochs
 were determined with a single model for each star as given in Table~\ref{tab:t06}.
For star 3589, we used the separately derived models given in Table~\ref{tab:t06}
 to derive abundances for each epoch before forming the weighted average.
Of course, M80 was observed during only the second epoch and the results were derived
 with the models listed in Table~\ref{tab:t06}.
The iron abundances assume a solar value of log~${\epsilon}~=~7.52$, and the rest
 of the abundances are on the \citet{AG89} scale.
The aluminum and lanthanum abundances are not corrected for hyperfine splitting (hfs),
 but the europium data are.
The hfs line list for the ${\lambda}6645$~{\AA} line was provided to us by
 C. Sneden (2002, private communication) and we used the blends driver in
 MOOG to determine the abundances.
The aluminum abundances are derived from only ${\lambda}6696$~{\AA}, since this
 feature is stronger than the other half of the doublet, and is often visible
 even when ${\lambda}6698$~{\AA} is not.
Most [Al/Fe] values were determined from equivalent width data, except for
 NGC~6752~--~3011 and M80~--~5, for which we fit synthetic spectra as required
 by the very weak lines.
All other abundances were determined via the equivalent width force-fitting routine
 in MOOG.

\placetable{tab:t08}

\begin{figure}[htb]
\epsscale{0.5}
\plotone{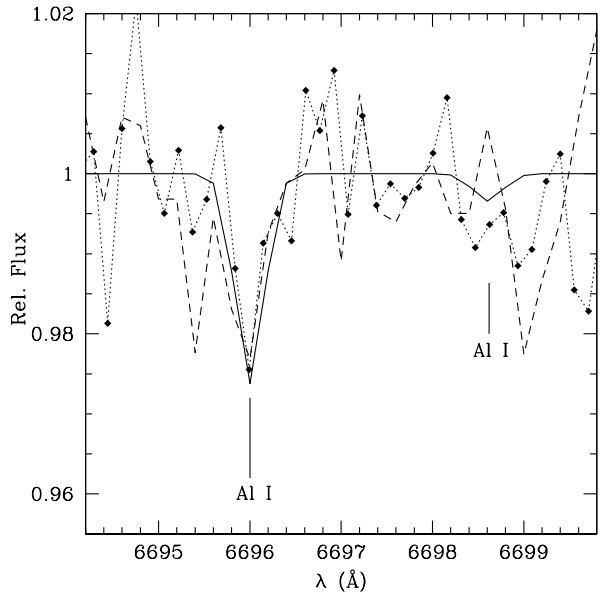}
\caption{Example of output from a Monte Carlo routine that adds noise (dashed
         line) to a synthetic spectrum (heavy solid line).  The data are shown
	 by filled diamonds connected by a light dotted line.}
\label{fig:f02}
\end{figure}

For lines that are indistinguishable from noise, we include estimates for upper 
 limits to [Al/Fe] for several stars in Table~\ref{tab:t08}.
These limits are found by fitting a synthetic spectrum to the data, then
 adding noise to the synthetic fit with an iterative Monte Carlo routine that
 outputs a noisy spectrum with a S/N ratio that is consistent with
 the actual data.
We iterate until the S/N ratio of the noisy synthetitc spectrum is
 within 5\% of the S/N of the data.
The synthetic spectrum is also sampled with the same spacing as the
 data and the line width is set by fitting a Gaussian to the Al~I
 line at ${\lambda}$6696 in a spectrum where it is easily measured.
The output is then compared by eye to the actual data.
This approach alleviates the challenge of trying to decide when a fit of
 a perfect synthetic spectrum to a real, noisy spectrum is actually
 measuring data or measuring noise.
All syntheses were carried out for the 2001 data, which had both high signal
 and superior resolution.
In Figure~\ref{fig:f02}, we show an example of the output from our routine for star
 NGC~6752~--~2403.
The S/N ratio of the data in this region is 124, versus
 120 in the ``noisy'' synthesis.
The results appear to indicate that the data might be a marginal
 detection of a weak Al~I feature at ${\lambda}$6696; however, we choose to err on the
 side of caution and consider this an upper limit.

The motive for deriving upper limits is to determine whether or not
 these stars might still be Al rich or whether they actually are Al normal
 and the Al lines are undetectable because of temperature,
 S/N ratio, or resolution.
The three stars in Table~\ref{tab:t08}
 with upper limits are indeed Al-normal to Al-poor.

\subsection{Discussion}
\label{sec:discuss}
\subsubsection{Aluminum}
\label{sec:aluminum}
\begin{figure}[htb]
\epsscale{0.5}
\plotone{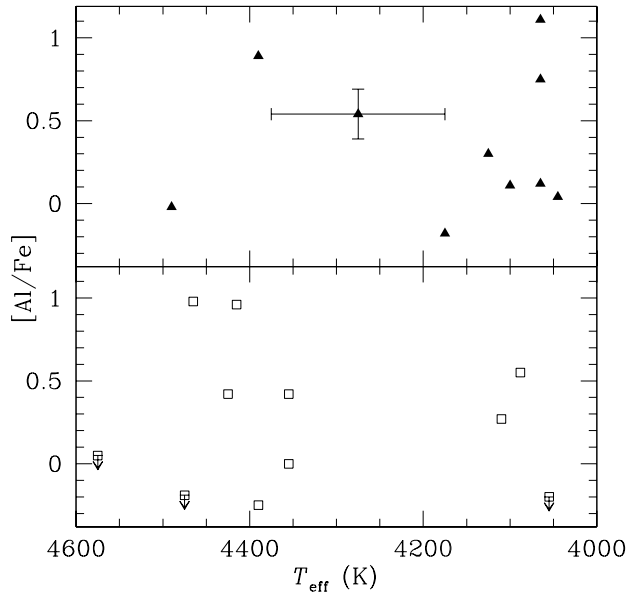}
\caption{[Al/Fe] as a function of {\Teff} for both clusters.  M80 data
         are denoted by filled triangles in the top panel and NGC~6752 data
	 by empty squares in bottom panel.  The upper limits for the NGC~6752 
	 data from Table~\ref{tab:t08} are shown as empty squares with
	 downward pointing arrows. The error bars shown in the top panel
	 are representative of all the data presented.}
\label{fig:f03}
\end{figure}

\begin{figure}[htb]
\epsscale{0.5}
\plotone{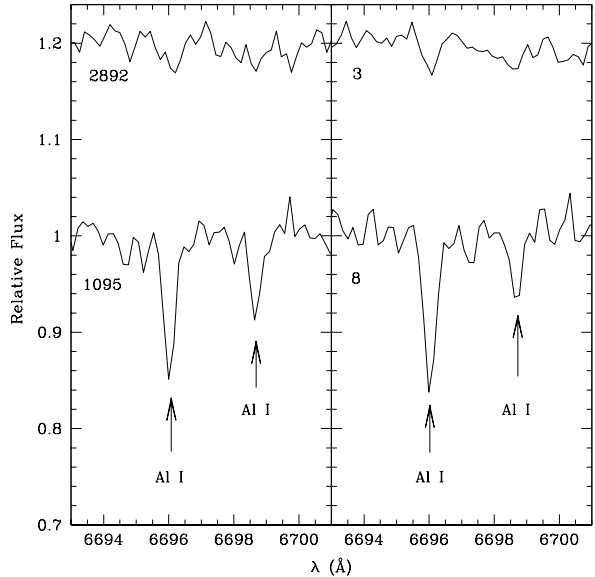}
\caption{Two spectra each from NGC~6752 (left panel) and M80 (right panel) that
         show strong variations in their Al~I features.  The spectra for
	 each cluster are taken from stars with similar effective temperatures.
	 The fluxes have been shifted for clarity.}
\label{fig:f04}
\end{figure}

The intent of this research is to discover the nature of the
 [Al/Fe] distribution in these two clusters, and
 we can make some observations despite the limited sample size.
Figure~\ref{fig:f03}, which shows [Al/Fe] as a function
 of the {\Teff}, demonstrates that there is a broad distribution
 along the RGB with no discernible trends in the Al ratio for either cluster.
As shown in Table~\ref{means}, which summarizes the mean
 abundances by cluster, aluminum is enhanced in both
 clusters and shows the largest distribution among the
 elements studied, suggesting that the abundance spread is
 not statistical.
That the variations are real is further supported in 
 Figure~\ref{fig:f04}, which shows the spectra of two stars
 with very similar atmospheric parameters from each 
 cluster having largely varying Al~I lines.
Our results are consistent with the giants studied by \citet{ND95}
 and \citet{GBNF2002}, and with the subgiants of \citet{GratAl2001}.
It would be useful to know how this distribution changes with
 magnitude by using the data of \citet{GratAl2001} and
 \citet{GBNF2002}, but theirs were purposely chosen in a biased manner
 according to $c_{1}$ indices and are understandably incomplete, given the 
 large number of stars in the magnitude ranges where they were operating,
 so that the data are insufficient to reveal any evolution of the Al ratio.
If our results are representative of the clusters' [Al/Fe] distributions in
 general, then one would have to rule out the connection between the Al
 ratio and the HB morphology, since we expect any ongoing mixing to
 create an upward trend in [Al/Fe] with decreasing {\Teff}, which is
 not seen here for this small sample.
Whether these [Al/Fe] distributions truly represent the actual distributions
 in these clusters needs to be demonstrated with larger
 datasets.


\subsubsection{Other elements}
\label{sec:other}
We look briefly at the other elements that had lines present in
 our spectra.
Consistent with other clusters, calcium, an ${\alpha}$-element,
 is enhanced by ${\sim}$0.25 dex relative to solar for
 both clusters.
On the other hand, titanium is difficult to interpret;
 the neutral lines do not always agree well with
 the results from the ionized lines.
On average the [Ti/Fe]$_{\rm II}$ abundances are higher
 than the neutral-line abundances, possibly indicating
 miscalculated gravities, NLTE effects, poor oscillator strengths, or
 a combination of any of these.
With regard to the gravities, we show in Table~\ref{tab:t07}
 that even a 0.2~dex change in {\logg} results in less
 than a 0.1~dex change in [Ti/Fe] as determined from the
 Ti~II lines, indicating that the discrepancy from the various
 lines probably has some other source.
The Fe-peak elements chromium and nickel track iron with
 no unusual trends.

In both clusters the abundance of the neutron-capture element Eu
is significantly enhanced relative to Fe, particularly
 for M80, while [La/Fe] is somewhat less enhanced for M80, and
 not at all for NGC~6752; however, the ratio of [Eu/La] remains 
 constant at $+0.42$.
This number is consistent with the results of \citet{Ivans2001}
 for M5, where the mean [Eu/La] is $+0.40$.
We do note that a straightforward comparison of lanthanum and
 europium abundances with the work of other authors is inhibited 
 by the fact that different authors use different atomic data,
 particularly when correcting for hfs effects, where the data
 are scarce and change frequently.
Still, the mean [Eu/Fe] for M80 is higher than seen in M13, which
 has $<[\rm Eu/Fe]>~=~+0.44~{\pm}~0.11$~({\stdv}) \citep{S96a}, or
 in either NGC~288 or NGC~362, which are enhanced by {\aprox}0.55 dex
 \citep{SK2000}.
In both papers the authors use a {\lgf} of $+0.20$, which if put
 on our scale would cause their abundances to decrease by 0.22 dex,
 making the discrepancy even larger.

Finally, we note via the ${\lambda}$6707~{\AA} Li~I feature
 that no cluster stars showed evidence for having
 enhanced lithium, as observed by \citet{Li} in M3, and
 is consistent with the results of \citet{PSKHW2000}.

\section{CONCLUSIONS}
\label{sec:conclude}
\subsection{Summary}
\label{sec:summary}

We begin the conclusions by summarizing our results:
\begin{itemize}
  \item We observed 21 giants in the globular clusters M80 and
        NGC~6752 with spectra of sufficient quality to determine
        abundances.
  \item Both M80 and NGC 6752 display a spread of aluminum abundances,
        having mean abundances
	of $+0.37$~dex and $+0.51$~dex, respectively.  The abundance
	spreads are 0.43 ({\stdv}) dex for M80 and 0.36 ({\stdv}) dex
	for NGC~6752.
  \item The aluminum data cannot resolve the discrepancy in the [Al/Fe]
        values near the main-sequence turnoff and the subgiants as observed by 
	\citet{GratAl2001}.
  \item No trends between [Al/Fe] ratios and magnitude are
        discernible in our small sample.
  \item The mean [Fe/H] value for M80 is $-1.73$, which is the first
        spectroscopic determination for this cluster.  For NGC~6752, the
	mean [Fe/H] is $-1.58$, which is consistent with previous results.
  \item The Fe-peak elements chromium and nickel follow iron closely.
  \item The [Ca/Fe] enhancements are consistent with the ${\alpha}$-enhancements
        observed in other clusters.
  \item The [Eu/La] ratio is constant for both clusters at $+0.42$ dex, which
	does not appear unusual; however,
        both [Eu/Fe] and [La/Fe] are enhanced in M80 relative to NGC~6752.
\end{itemize}

\subsection{A Final Look At Aluminum}
\label{sec:final_look}

Now that we've derived aluminum abundances in M80 and NGC~6752,
 are we any closer to answering whether or not the variations
 are the result of primordial pollution, mixing processes, or
 both?
Perhaps when combined with the results of
 \citet{GBNF2002} we can get some insight into just how complicated
 this problem really is (we do not include the Gratton data here since
 they were biased by use of the $c_{1}$ index to select stars.)
Given the differences in the quality of the data between the two studies, and
 the still small number of stars that have been analyzed,
 it is difficult to draw firm conclusions.
If mixing is an ongoing phenomenon, then one would expect that the
 Al ratio would increase with decreasing magnitude, as it appears
 to in M13 \citep{CN2000}; unfortunately,
 the small numbers make this difficult to discern at this point.

An interesting result also comes from both \citet{GratAl2001}
 and \citet{GBNF2002}, who show that stars on the subgiant/lower red giant
 branch may have just as much Al in them as stars on the upper RGB,
 where mixing is theoretically possible.
These results might suggest that all Al anomalies are the
 result of pollution, but this raises yet another question:
 Can it be demonstrated that a globular cluster main sequence star,
 subgiant, or lower red giant exists that has strongly enhanced Al
 and is not depleted in C and/or O?
That is, if the CNO variations are the result of deep mixing and
 the Al (and Na) anomalies are primordial, there should be stars
 that show uncorrelated abundance patterns.
But, if the CNO and Na-Al anomalies are both caused by primordial
 scenarios, then why does evidence exist, in clusters where it's
 been studied, that the total C$+$N$+$O remains constant from
 star to star independent of the Na or Al content \citep{NBS81,
 CAP88,DCCB91}, and how does one explain $^{12/13}$C ratios near the CN cycle
 equilibrium value in many other clusters \citep{SS91,S96b,ZWB96,
 BSKL97,BSSBN97}?
On the other hand, if the anomalies are all created {\em in-situ},
 what is the physics behind them that can occur without completely
 contradicting well-established basic theories of stellar evolution?
It used to be that one could side with one scenario or the other,
 then the data became more complicated, and one would say that
 it was somehow a mixture of both primordial and evolutionary
 scenarios.
Maybe in the end that will be the final conclusion, but we are far from proving
 it; too many questions remain to be answered and too many observations remain to
 be made.

\acknowledgements

We thank D. Dinescu for providing us with valuable proper motion
 data, P. Palunas for aiding us with coordinate system transformations, and
 S. Keller for helping us with our photometry.
We also wish to thank C. Sneden for providing us with both useful atomic
 data and advice on running MOOG.
Finally, we thank our anonymous referee for his/her careful reading of
 our manuscript and helpful comments.
In addition, R. M. C. acknowledges the National Research Council for support
 during a part of this work, as well as the AAS Small Research Grant program,
 and the Djehuty Stellar Evolution Project.
This work was performed under the auspices of the U.S. Department of Energy,
National Nuclear Security Administration by the University of California,
Lawrence Livermore National Laboratory under contract No. W-7405-Eng-48.
C.A.P. gratefully acknowledges support from the Daniel Kirkwood Research Fund at
Indiana University.
This research has made use of the SIMBAD database, operated at CDS, Strasbourg, France.

\newpage
\begin{deluxetable}{lll}
\tablewidth{0pt}
\tablecolumns{3}
\tablecaption{Observing Conditions and Instrument Parameters \label{tab:t01}}
\tablehead{
       \colhead{} & \colhead{Epoch 1} & \colhead{Epoch 2}
 }
 
%===================================================================
\startdata
%                                                    Epoch 1               Epoch 2
Cluster(s)                                   & NGC 6752               & NGC 6752, M80\\
Dates (UT)                                   & 26 June 1999           & 22-23 June 2001\\
Fibers                                       & Large\tablenotemark{a} & Large\tablenotemark{a}\\
Slit Plates (${\mu}$)                        & None                   & 200\tablenotemark{b} \\
$f_{\rm cam.}$ (mm)                          & 229                    & 400 \\
CCD                                          & Loral 1k${\times}$3k   & SiTe 2k${\times}$4k \\
Binning                                      & 1 ${\times}$ 1         & 1 ${\times}$ 2\\
Central Wavelength ({\AA})                   & 6725                   & 6667 \\
Resolution                                   & 9400                   & 18000\\
${\Delta}{\lambda}$\tablenotemark{c} ({\AA}) & 485                    & 300 \\
Seeing ({\arcsec})                           & 1.4                    & 0.74 \\
Moon Age (days from new)                     & 13                     & 1-2  
\enddata
%===================================================================
\tablenotetext{a}{300 ${\mu}$ (2{\arcsec}) fibers}
\tablenotetext{b}{Projects to 1{\farcs}3}
\tablenotetext{c}{Refers to the width of the spectrum on the CCD}
\end{deluxetable}

\newpage
\begin{deluxetable}{lrccclll}
\tablewidth{0pt}
\tablecolumns{8}
\tablecaption{Stars Observed in NGC 6752 \label{tab:t02}}
\tablehead{
  \colhead{Star} & \colhead{Alt.} & \colhead{V}     & \colhead{($B-V$)$_{0}$} & \colhead{t$_{exp}$ (s)} & \colhead{S/N}       & \colhead{t$_{exp}$ (s)} & \colhead{S/N} \\
  \colhead{(1)} & \colhead{(2)} & \colhead{(1)} & \colhead{(1)}       & \colhead{(Epoch 1)}     & \colhead{(Epoch 1)} & \colhead{(Epoch 2)}     & \colhead{(Epoch 2)}
 }
%===================================================================
% Bun  Din    V      B-Vo  T_exp(1) S/N(1) Texp(2) S/N(2)
\startdata
3589 & 231 & 10.85 & 1.73 & 9600  & 230 & 7200 & 130 \\
2403 & 242 & 10.95 & 1.43 & 9600  & 190 & 7200 & 125 \\
2113 & 244 & 11.23 & 1.37 & 9600  & 190 & 7200 & 135 \\
277  & 379 & 11.43 & 1.13 & 9600  & 185 & 7200 & 135 \\
2240 & 590 & 11.63 & 1.13 & 9600  & 275 & 7200 & 120 \\
1518 & 250 & 11.82 & 1.07 & 16800 & 245 & 7200 &  75 \\
3011 & 239 & 11.98 & 1.10 & 9600  & 135 & 7200 &  70 \\
3805 & 136 & 12.00 & 1.08 & 9600  & 135 & 7200 &  95 \\
1095 & 598 & 12.18 & 1.04 & 9600  & 130 & 7200 &  70 \\
2892 & 268 & 12.22 & 1.03 & 16800 & 140 & 7200 &  90 \\
4437 & 93  & 12.78 & 0.96 & 16800 & 100 & 7200 &  70   
\enddata
%===================================================================
\tablecomments{$V$ and {\bv} are from \citet{ngc6752}, using their
 value of E($B~-~V)~=~0.04$ to correct the colors. Epochs 1 and 2 are described
 in Table~\ref{tab:t01}.}
\tablerefs{1 \citep{ngc6752}, 2 (D. Dinescu 1998, private communication}
\end{deluxetable}

\newpage
\begin{deluxetable}{lllll}
\tablewidth{0pt}
\tablecolumns{5}
\tablecaption{Stars Observed in M80 \label{tab:t03}}
\tablehead{
  \colhead{Star} & \colhead{V} & \colhead{(B-V)$_{0}$} & \colhead{t$_{exp}$ (s)} & \colhead{S/N}
 }
%===================================================================
\startdata
1   & 12.83 & 1.44 & 13800 & 185 \\
2   & 12.96 & 1.60 & 13800 & 120 \\
3   & 12.97 & 1.42 & 13800 & 200 \\
4   & 12.98 & 1.38 & 13800 & 130 \\
5   & 12.99 & 1.30 & 13800 &  70 \\
8   & 13.28 & 1.42 & 13800 &  60 \\
11  & 13.40 & 1.20 & 13800 & 100 \\
16  & 13.45 & 1.42 & 13800 &  90 \\
23  & 13.72 & 1.02 & 13800 & 100 \\ 
31  & 13.99 & 1.10 & 13800 & 110   
\enddata
%===================================================================
\end{deluxetable}

\newpage
\begin{deluxetable}{lllllllllllllll}
\tablewidth{0pt}
\rotate
\tablecolumns{15}
\tablecaption{Atomic Line Parameters and Equivalent Widths for NGC 6752 Giants \label{tab:t04}}
\tablehead{
 \colhead{${\lambda}$} & \colhead{Species} & \colhead{E.P.} & \colhead{Log gf} & \colhead{3589} & \colhead{2403} & \colhead{2113} & \colhead{277} & \colhead{2240} & \colhead{1518} & \colhead{3011} & \colhead{3805} &  \colhead{1095} & \colhead{2892} & \colhead{4437} \\
  \colhead{({\AA})}     & \colhead{}        & \colhead{(eV)} & \multicolumn{12}{c}{}
 }
%=========================================================================================================================
\startdata
%=========================================================================================================================
% Wl        Sp     EP     log gf  3589    2403    2113   277      2240    1518   3011     3805   1095     2892    4437
6696.023 & Al I  & 3.14 & $-$1.35 & 57.5  & \nd   & 28.7  & 23.8  & 13.0  & 35.9  &  6.5  & 67.9  & 59.0  & \nd   & \nd   \\
         &       &      &         & 55.3  & \nd   & \nd   & 25.6  & 17.4  & 25.9  & \nd   & 54.4  & 56.5  & \nd   & \nd   \\
6698.673 & Al I  & 3.14 & $-$1.65 & 30.8  & \nd   & 12.3  & \nd   & \nd   & \nd   & \nd   & 28.7  & 23.3  & \nd   & \nd   \\
         &       &      &         & 33.2  & \nd   & \nd   & \nd   & \nd   & \nd   & \nd   & 26.8  & 31.5  & \nd   & \nd   \\
6717.687 & Ca I  & 2.71 & $-$0.39 &  \nd  & \nd   & \nd   & \nd   & \nd   & \nd   & \nd   & \nd   & \nd   & \nd   & \nd   \\
         &       &      &         & 133.1 & 107.1 & 127.3 & 76.7  & 109.2 & 93.8  & 84.3  & 88.9  & 93.8  & 93.0  & 72.2  \\
6599.113 & Ti I  & 0.90 & $-$2.06 & 118.0 & 22.1  & 57.6  & 16.4  & 32.2  & 12.2  & \nd   & \nd   & \nd   & \nd   & \nd   \\
         &       &      &         & 94.6  & 27.4  & 63.6  & 18.4  & 37.6  & \nd   & \nd   & 13.6  & \nd   & 20.5  & \nd   \\
6743.127 & Ti I  & 0.90 & $-$1.76 & 126.9 & 42.4  & 74.5  & \nd   & 51.4  & 33.4  & 27.7  & \nd   &  \nd  & 17.9  & \nd   \\
         &       &      &         & 112.1 & \nd   & 89.2  & 31.1  & 56.9  & 32.8  & 36.3  & 37.2  & 19.5  & \nd   & \nd   \\
6606.979 & Ti II & 2.06 & $-$2.85 &   \nd &  \nd  & 27.2  &  \nd  & 19.7  & \nd   & \nd   & \nd6  & \nd   & \nd   & \nd   \\
         &       &      &         & 27.4  & 21.9  & 35.6  & 17.8  & 19.9  & \nd   & \nd   & 21.6  & \nd   & \nd   & \nd   \\
6630.032 & Cr I  & 1.03 & $-$3.59 & 39.0  & \nd   & 22.9  & \nd   &  \nd  & \nd   & \nd   & \nd   & \nd   & \nd   & \nd   \\
         &       &      &         & 44.1  & \nd   & \nd   & \nd   & 11.9  & \nd   & \nd   & \nd   & \nd   & \nd   & \nd   \\
6581.210 & Fe I  & 1.49 & $-$4.82 & \nd   & \nd   & \nd   & \nd   &  \nd  & \nd   & \nd   & \nd   & \nd   & \nd   & \nd   \\
         &       &      &         & 89.0  & 55.2  & 70.2  & 30.9  &  \nd  & 47.9  & \nd   & 40.6  & 38.3  & 34.5  & \nd   \\
6593.870 & Fe I  & 2.43 & $-$2.34 & 150.3 & 125.4 & 137.3 & 108.5 & 108.1 & 82.7  & 106.1 & 107.6 & 93.3  & 97.8  & \nd   \\
         &       &      &         & 147.8 & 129.0 & 151.0 & 115.5 & 106.5 & 116.5 & 95.8  & 96.7  & 89.7  & 98.6  & 80.1  \\
6608.026 & Fe I  & 2.28 & $-$4.02 & 50.3  &  \nd  & 40.7  &  \nd  & 24.6  &  \nd  & \nd   & \nd   &  \nd  & \nd   & \nd   \\
         &       &      &         & 52.4  & 27.2  & 43.9  & 15.3  & 25.4  & 20.6  & 20.0  &  \nd  & 20.2  & \nd   & 10.4  \\
6609.110 & Fe I  & 2.56 & $-$2.67 & \nd   & \nd   & \nd   & \nd   &  \nd  & \nd   & \nd   & \nd   & \nd   & \nd   & \nd   \\
         &       &      &         & 116.4 & 85.1  & 106.1 & 81.3  & 85.7  & 84.7  & 62.6  & 74.9  & 65.3  & 66.0  & 58.2  \\
6609.679 & Fe I  & 0.99 & $-$5.87 & \nd   & \nd   & \nd   & \nd   &  \nd  & \nd   & \nd   & \nd   & \nd   & \nd   & \nd   \\
         &       &      &         & 61.9  & 36.6  & 48.9  & 18.4  & 27.3  & 18.6  & 11.5  & 23.8  & 13.1  & 12.4  & 10.8  \\
6625.022 & Fe I  & 1.01 & $-$5.38 & 122.2 & 59.5  & 81.3  & 43.0  & 55.5  & 33.5  & 28.0  & \nd   &  \nd  & 32.8  & \nd   \\
         &       &      &         & 110.6 & 54.8  & 85.8  & 38.5  & 50.2  & 43.6  & 34.8  & 38.9  & 29.9  & 42.0  & 23.7  \\
6646.932 & Fe I  & 2.61 & $-$4.01 & 27.2  &  7.8  &  \nd  &  \nd  & 13.7  & 11.2  &  \nd  & \nd   & \nd   &  \nd  & \nd   \\
         &       &      &         & 26.3  & \nd   & 16.9  & 7.4   & 13.3  & 19.0  & 6.0   & \nd   & \nd   & 10.9  & \nd   \\
6648.081 & Fe I  & 1.01 & $-$5.88 & 53.1  & 18.9  & 38.5  &  9.1  & 33.1  & 15.9  &  \nd  &  \nd  & \nd   & 14.1  & \nd   \\
         &       &      &         & 58.6  & 15.7  & 35.0  & 12.6  & 24.8  & 21.3  & 13.3  & 13.8  & \nd   & 15.4  & \nd   \\
6667.419 & Fe I  & 2.45 & $-$4.42 & 18.8  &  \nd  &  \nd  & \nd   & \nd   & \nd   & \nd   & \nd   & \nd   & \nd   & \nd   \\
         &       &      &         & 19.0  & \nd   & 20.6  & 7.3   & \nd   & \nd   & \nd   & \nd   & \nd   & \nd   & \nd   \\
6677.987 & Fe I  & 2.69 & $-$1.22 & 219.5 & 171.0 & 170.4 & 150.7 & 154.2 & 142.9 & 131.2 & 142.2 & \nd   & 126.1 & \nd   \\
         &       &      &         & 182.3 & 165.1 & 172.5 & 154.0 & 156.8 & 143.8 & 140.7 & 136.1 & 135.2 & 128.2 & 109.3 \\
6703.567 & Fe I  & 2.76 & $-$3.13 & 65.1  & 48.0  & 45.0  & 34.1  & 39.8  & 32.3  & 24.6  & 37.7  & 28.5  & 20.3  & \nd   \\
         &       &      &         & 68.0  & 51.4  & 60.2  & 29.4  & 46.6  & 37.8  & 30.8  & 38.0  & 33.3  & 31.7  & 24.8  \\
6710.320 & Fe I  & 1.49 & $-$4.90 & 82.2  & 36.5  & 55.7  & 35.0  & 43.7  & 32.8  & 37.6  & 28.3  & 26.5  & 25.4  & \nd   \\
         &       &      &         & 74.7  & \nd   & 57.0  & 30.6  & 45.8  & 33.4  &  \nd  & 22.4  & 29.1  & 17.9  & 22.0  \\
6739.522 & Fe I  & 1.56 & $-$4.98 & 89.9  & 21.4  & 54.9  & 42.4  & 45.8  & 28.7  &  \nd  & 27.0  &  \nd  & 14.0  & \nd   \\
         &       &      &         & 62.0  & \nd   & 42.5  & 22.0  & 29.3  & 22.2  & 19.8  & 15.9  & 22.6  & 21.1  & 12.4  \\
6750.153 & Fe I  & 2.42 & $-$2.48 & 134.1 & 113.9 & 120.4 & 80.0  & 102.1 & 82.1  & 84.9  &  \nd  & 89.9  & 83.2  & \nd   \\
         &       &      &         & 131.6 & 104.6 & 117.3 & 98.7  & 103.3 & 86.3  & 99.5  & 90.2  & 78.8  & 90.8  & 65.9  \\
6752.707 & Fe I  & 4.64 & $-$1.30 & 47.2  & \nd   & 23.2  &  \nd  & 24.2  & \nd   & 10.4  & 11.6  & 11.3  &  \nd  & \nd   \\
         &       &      &         & 45.2  & \nd   & 20.2  & 5.4   & 16.8  & \nd   & 11.8  & 9.9   & \nd   & 14.7  & 11.2  \\
6806.845 & Fe I  & 2.73 & $-$3.24 & 68.2  & 33.8  & 42.6  & 33.6  & 43.3  & 30.5  & 28.8  & 38.3  &  \nd  & 25.5  & \nd   \\
         &       &      &         & \nd   & \nd   & \nd   & \nd   & \nd   & \nd   & \nd   & \nd   & \nd   & \nd   & \nd   \\
6810.263 & Fe I  & 4.61 & $-$1.12 & 27.0  & 13.0  & 23.2  & \nd   & 23.5  & 17.1  & \nd   & \nd   & \nd   & 11.2  & \nd   \\
         &       &      &         & \nd   & \nd   & \nd   & \nd   & \nd   & \nd   & \nd   & \nd   & \nd   & \nd   & \nd   \\
6839.831 & Fe I  & 2.56 & $-$3.47 & 65.2  & 30.0  & 47.8  & \nd   & 28.5  & 22.3  & \nd   & \nd   & \nd   & \nd   & \nd   \\
         &       &      &         & \nd   & \nd   & \nd   & \nd   & \nd   & \nd   & \nd   & \nd   & \nd   & \nd   & \nd   \\
6586.319 & Ni I  & 1.95 & $-$2.95 & \nd   & \nd   & \nd   & \nd   &  \nd  & \nd   & \nd   & \nd   & \nd   & \nd   & \nd   \\
         &       &      &         & 79.7  & 64.9  & 65.1  & 46.6  & 57.7  & 42.4  & 59.3  & 41.2  & 38.3  & 34.9  & \nd   \\
6643.638 & Ni I  & 1.68 & $-$2.01 & 176.4 & 129.8 & 145.3 & 122.4 & 127.6 & 111.0 & 109.7 & 114.5 & 92.9  & 86.6  & \nd   \\
         &       &      &         & 171.6 & 142.5 & 145.5 & 123.4 & 128.3 & 125.4 & 112.4 & 108.7 & 106.0 & 94.3  & 77.7  \\
6767.784 & Ni I  & 1.83 & $-$1.89 & 137.0 & 121.7 & 113.1 & 107.7 & 102.9 & 89.9  &  87.4 & 88.8  & 97.4  & 89.0  & \nd   \\
         &       &      &         & 141.4 & \nd   & 124.1 & 116.0 & 106.6 & 83.7  & 106.4 & 99.9  & 90.9  & 85.0  & 77.7  \\
6772.321 & Ni I  & 3.66 & $-$1.07 & 31.7  & 31.7  & 29.9  & 23.3  & 28.4  & 21.6  & 17.6  & 16.9  & 26.6  & 15.5  & \nd   \\
         &       &      &         & 39.1  & \nd   & 43.7  & 30.9  & 29.9  & 41.5  & 25.6  & 27.9  & 25.3  & \nd   & 19.7  \\
6774.330 & La II & 0.13 & $-$1.75 & 35.1  & 17.5  & 24.3  & \nd   & 16.1  &  9.3  & 12.0  &  8.1  & \nd   & \nd   & \nd   \\
         &       &      &         & 31.4  & \nd   & 28.8  & \nd   & 14.4  & 14.0  & \nd   & \nd   & \nd   & \nd   & \nd   \\
6645.127 & Eu II & 1.38 & $+$0.42 & 32.1  & 25.2  & 30.4  & 25.2  & 28.6  & 23.4  & 23.7  & \nd   & 14.4  & \nd   & \nd   \\
         &       &      &         & 30.7  & 13.3  & 26.7  & 19.9  & 24.6  & 27.7  & 23.3  & 16.5  & \nd   & 13.6  & \nd     
% Wl        Sp     EP     log gf  3589    2403    2113    277     2240    1518    3011    3805    1095    2892    4437
%==========================================================================================================================
\enddata
\tablecomments{The first row of each line refers to the 1999 data, while the second row refers to the 2001 data.}
\end{deluxetable}

\newpage
\begin{deluxetable}{llllllllllllll}
\tablewidth{0pt}
\rotate
\tablecolumns{14}
\tablecaption{Atomic Line Parameters and Equivalent Widths for M80 Giants \label{tab:t05}}
\tablehead{
 \colhead{${\lambda}$} & \colhead{Species} & \colhead{E.P.} & \colhead{Log gf} & \colhead{1} & \colhead{2} & \colhead{3} & \colhead{4} & \colhead{5} & \colhead{8} & \colhead{11} & \colhead{16} & \colhead{23} & \colhead{31} \\
 \colhead{({\AA})}     & \colhead{}        & \colhead{(eV)} & \multicolumn{11}{c}{}
 }
\startdata
%================================================================================================================================================
% wl        El       EP       log gf     1       2       3       4       5       8       11      16      23      31
 6696.023 & Al I  &   3.14  &  -1.35 &  16.1  & 39.5  & 16.2  & 16.3  & \nd   & 72.8  & 38.4  & 38.7  & 7.8   & 55.4  \\
 6698.673 & Al I  &   3.14  &  -1.65 &  \nd   & 17.8  & \nd   & \nd   & \nd   & 27.6  & 13.5  & 26.3  & \nd   & 27.8  \\
 6717.687 & Ca I  &   2.71  &  -0.39 &  109.4 & 131.8 & 108.3 & 104.2 & 118.2 & 119.1 & 82.1  & 99.9  & 72.4  & 82.6  \\
 6599.113 & Ti I  &   0.90  &  -2.06 &  49.6  & 93.0  & 38.1  & 52.6  & 52.6  & 33.1  & \nd   & \nd   & \nd   & \nd   \\
 6743.127 & Ti I  &   0.90  &  -1.76 &  67.8  & 122.1 & 71.0  & 65.4  & 69.8   & 53.2 & 41.4  & 29.6  & \nd   & 17.5  \\
 6606.979 & Ti II &   2.06  &  -2.85 &  14.4  & 19.8  & 20.7  & 18.3  & 29.3  & \nd   & \nd   & \nd   & \nd   & 17.2  \\
 6630.032 & Cr I  &   1.03  &  -3.59 &  18.2  & 47.6  & 18.2  & 21.9  & \nd   & \nd   & \nd   & \nd   & \nd   & \nd   \\
 6581.210 & Fe I  &   1.49  &  -4.82 &  63.3  & 83.9  & 55.5  & 50.3  & 62.4  & 59.8  & 46.1  & 46.6  & 16.7  & 33.2  \\
 6593.871 & Fe I  &   2.43  &  -2.34 &  127.2 & 162.4 & 126.6 & 147.7 & 113.4 & 122.8 & 101.0 & 123.5 & 73.0  & 96.2  \\
 6608.026 & Fe I  &   2.28  &  -4.02 &  31.6  & 39.6  & 29.1  & 122.6 & 38.2  & 27.4  & 22.2  & \nd   & \nd   & \nd   \\
 6609.110 & Fe I  &   2.56  &  -2.67 &  91.1  & 114.2 & 97.2  & 30.2  & 85.5  & 90.8  & 82.1  & 77.4  & 62.7  & 66.7  \\
 6609.679 & Fe I  &   0.99  &  -5.87 &  44.4  & 52.2  & 34.5  & 97.0  & 35.3  & 27.3  & 30.9  & 40.0  & 6.8   & 16.7  \\
 6625.022 & Fe I  &   1.01  &  -5.38 &  73.6  & 106.6 & 70.8  & 32.4  & 60.4  & 62.9  & 47.1  & 44.9  & 21.8  & 30.5  \\
 6646.932 & Fe I  &   2.61  &  -4.02 &  15.1  & 17.7  & 12.0  & 68.1  & 12.4  & \nd   & 12.6  & 11.1  & \nd   & \nd   \\
 6648.081 & Fe I  &   1.01  &  -5.88 &  35.0  & 49.9  & 29.5  & 14.3  & 20.8  & 27.5  & 12.1  & 34.0  & 18.0  & 17.3  \\
 6667.419 & Fe I  &   2.45  &  -4.42 &  14.2  & 24.2  & 12.8  & 24.4  & 10.9  & \nd   & \nd   & \nd   & \nd   & 4.4   \\
 6677.987 & Fe I  &   2.69  &  -1.22 &  170.6 & 192.8 & 167.0 & \nd   & 161.0 & 163.5 & 141.6 & 153.0 & 120.2 & \nd   \\
 6703.567 & Fe I  &   2.76  &  -3.13 &  59.4  & 59.4  & 48.3  & 50.1  & 50.9  & 45.5  & 32.1  & 28.9  & 29.2  & \nd   \\
 6710.320 & Fe I  &   1.49  &  -4.90 &  55.3  & 63.9  & 43.8  & 51.8  & 68.1  & 64.8  & 23.6  & 29.7  & \nd   & 23.6  \\
 6739.522 & Fe I  &   1.56  &  -4.98 &  40.5  & 36.5  & 33.2  & 35.0  & 22.9  & \nd   & 28.5  & 29.8  & \nd   & 14.9  \\
 6750.153 & Fe I  &   2.42  &  -2.48 &  119.2 & 132.8 & 113.8 & 112.0 & 128.5 & 98.3  & 100.9 & 104.8 & 88.3  & 84.9  \\
 6752.707 & Fe I  &   4.64  &  -1.30 &  16.3  & 27.0  & \nd   & 11.8  & \nd   & 12.6  & \nd   & 13.4  & \nd   & 13.0  \\
 6586.319 & Ni I  &   1.95  &  -2.95 &  67.9  & 77.7  & 69.1  & 60.9  & 75.0  & 54.5  & 51.6  & 54.0  & 30.7  & 38.8  \\
 6643.638 & Ni I  &   1.68  &  -2.01 &  147.9 & 173.9 & 142.6 & 139.8 & 157.4 & 122.0 & 114.3 & 133.6 & 98.3  & 97.8  \\
 6767.784 & Ni I  &   1.83  &  -1.89 &  127.0 & 171.2 & 116.4 & 128.0 & 138.5 & 119.2 & 103.5 & 123.5 & 87.6  & 92.1  \\
 6772.321 & Ni I  &   3.66  &  -1.07 &  32.3  & \nd   & 28.8  & 26.3  & 37.3  & 24.5  & 25.8  & 30.1  & 19.9  & 22.4  \\
 6774.330 & La II &   0.13  &  -1.75 & 30.5   & \nd   & 26.6  & 23.5  & 15.4  & 31.4  & 20.6  & 17.1  & \nd   & 23.3  \\
 6645.127 & Eu II &   1.38  &  +0.42 & 32.4   & 39.7  & 42.4  & 29.5  & 27.8  & 28.8  & 27.9  & 27.4  & 24.2  & 24.0    
\enddata
% wl        El       EP       log gf     1       2       3       4       5       8       11      16      23      31
%==========================================================================
\end{deluxetable}

\newpage
\begin{deluxetable}{lcccc}
\tablewidth{0pt}
\tablecolumns{5}
\tablecaption{Model Atmosphere Parameters \label{tab:t06}}
\tablehead{
 \colhead{Star} & \colhead{{\Teff}} & \colhead{Log g}         & \colhead{{\vt}}         & \colhead{P/S\tablenotemark{a}} \\
 \colhead{}     & \colhead{(K)}     & \colhead{(cm s$^{-2}$)} & \colhead{(km s$^{-1}$)} & \colhead{}
 }
%=======================================================================================================================
\startdata
\sidehead{NGC 6752}
3589\tablenotemark{b} & 4050 & 0.55 & 2.20 & S \\
     & 4125 & 0.70 & 1.77 &   \\
2403 & 4055 & 0.56 & 2.11 & P \\
2113 & 4110 & 0.66 & 2.07 & P \\
 277 & 4355 & 1.13 & 1.85 & P \\
2240 & 4355 & 1.13 & 1.85 & P \\
1518 & 4425 & 1.27 & 1.79 & P \\
3011 & 4390 & 1.20 & 1.82 & P \\
3805 & 4415 & 1.25 & 1.80 & P \\
1095 & 4465 & 1.35 & 1.76 & P \\
2892 & 4475 & 1.38 & 1.75 & P \\
4437 & 4575 & 1.58 & 1.67 & P \\
\sidehead{M80}
1    & 4045 & 0.60 & 2.12 & P \\
2    & 4125 & 0.70 & 2.21 & S \\
3    & 4065 & 0.57 & 2.10 & P \\
4    & 4100 & 0.65 & 2.07 & P \\
5    & 4175 & 0.79 & 2.01 & P \\
8    & 4065 & 0.57 & 2.10 & P \\
11   & 4275 & 0.98 & 1.92 & P \\
16   & 4065 & 0.57 & 2.10 & P \\
23   & 4490 & 1.20 & 1.74 & P \\
31   & 4390 & 1.28 & 1.82 & P 
\enddata
%=================================================================================
\tablenotetext{a}{P = preferred model from photometry; S = preferred model from spectroscopy}
\tablenotetext{b}{First row refers to model parameters derived from the first-epoch spectrum, while the
                  second row refers to model parameters derived from the second-epoch spectrum.}
\end{deluxetable}

\newpage
\begin{deluxetable}{lccc}
\tablewidth{0pt}
\tablecolumns{4}
\tablecaption{Abundance Sensitivity to Model Atmosphere Variations\label{tab:t07}}
\tablehead{
	\colhead{Abundance} & \colhead{${\Delta}${\Teff}} & \colhead{${\Delta}${\logg}} & \colhead{${\Delta}${\vt}} \\
	\colhead{Ratio}     & \colhead{${\pm}$100~K}     & \colhead{${\pm}$0.2~dex}   & \colhead{${\pm}$0.2~{\kms}}
}
\startdata
${\Delta}$[Fe/H]             & 0.15 & 0.02 & 0.04 \\
${\Delta}$[Al/Fe]            & 0.09 & 0.01 & 0.00 \\
${\Delta}$[Ca/Fe]            & 0.13 & 0.01 & 0.10 \\
${\Delta}$[Ti/Fe]$_{\rm I}$  & 0.21 & 0.01 & 0.03 \\
${\Delta}$[Ti/Fe]$_{\rm II}$ & 0.03 & 0.09 & 0.02 \\
${\Delta}$[Cr/Fe]            & 0.20 & 0.01 & 0.00 \\
${\Delta}$[Ni/Fe]            & 0.11 & 0.03 & 0.08 \\
${\Delta}$[La/Fe]            & 0.03 & 0.09 & 0.01 \\
${\Delta}$[Eu/Fe]            & 0.01 & 0.09 & 0.01 
\enddata
\end{deluxetable}

\newpage
\begin{deluxetable}{llllllllll}
\tablewidth{0pt}
\tabletypesize{\footnotesize}
\tablecolumns{10}
\tablecaption{Abundances\label{tab:t08}}
\tablehead{
            \colhead{Star} & \colhead{[Fe/H]} & \colhead{[Al/Fe]} & \colhead{[Ca/Fe]} & \colhead{[Ti/Fe]$_{\rm I}$} & \colhead{[Ti/Fe]$_{\rm II}$} & \colhead{[Cr/Fe]} & \colhead{[Ni/Fe]} & \colhead{[La/Fe]} & \colhead{[Eu/Fe]}
 }
%=======================================================================================================================
\startdata
\sidehead{NGC 6752}
%         Fe       Al         Ca     Ti I     Ti II    Cr          Ni      La        Eu      %How arrived:
3589 & $-$1.31 & +0.55    & +0.33 & +0.45 & +0.22 & $-$0.02   & $-$0.15 & +0.09   & +0.36 \\ %Average = (2x2001 + 1999)/3
2403 & $-$1.94 & $<-0.20$ & +0.24 & $-$0.04 & +0.71 & \nd     & +0.12   & +0.34   & +0.62 \\ %Average
2113 & $-$1.60 & +0.27    & +0.32 & +0.25   & +0.64 & $-$0.06 & $-$0.17 & +0.21   & +0.57 \\ %Average
277  & $-$1.66 & +0.42    & +0.11 & +0.09   & +0.48 & \nd     & +0.02   & \nd     & +0.66 \\ %Average
2240 & $-$1.47 & +0.00    & +0.39 & +0.28   & +0.34 & $-$0.06 & $-$0.15 & +0.08   & +0.57 \\ %Average
1518 & $-$1.48 & +0.42    & +0.20 & \nd     & +0.06 & \nd     & $-$0.17 & +0.07   & +0.64 \\ %Average
3011 & $-$1.69 & $-0.25$  & +0.25 & \nd     & +0.23 & \nd     &  0.00   & +0.24   & +0.76 \\ %Average
3805 & $-$1.57 & +0.96    & +0.25 & +0.13   & +0.54 & \nd     & $-$0.15 & $-$0.10 & +0.48 \\ %Average
1095 & $-$1.57 & +0.98    & +0.38 & $-$0.05 & \nd   & \nd     & $-$0.13 & \nd     & +0.45 \\ %Average
2892 & $-$1.55 & $<-0.19$ & +0.27 & +0.16   & \nd   & \nd     & $-$0.31 & \nd     & +0.41 \\ %Average
4437 & $-$1.58 & $<+0.05$ & +0.18 & \nd     & \nd   & \nd     & $-$0.24 & \nd     & \nd   \\ %Average
\sidehead{M80}
%      Fe         Al          Ca     Ti I    Ti II      Cr         Ni      La        Eu      %Model atm. from:
 1   & $-$1.72 & +0.04    & +0.17 & +0.09   & +0.35 & $-$0.19 & $-$0.04 & +0.44   & +0.78 \\ %Photometry
 2   & $-$1.47 & +0.30    & +0.27 & +0.53   & +0.22 & +0.20   & +0.03   & \nd     & +0.66 \\ %Spectra
 3   & $-$1.82 & +0.12    & +0.24 & +0.13   & +0.57 & $-$0.10 & $-$0.05 & +0.41   & +1.05 \\ %Photometry
 4   & $-$1.79 & +0.11    & +0.20 & +0.24   & +0.51 & +0.03   & $-$0.06 & +0.35   & +0.82 \\ %Photometry
 5   & $-$1.66 & $-0.18$  & +0.41 & +0.31   & +0.66 &  \nd    & +0.14   & +0.09   & +0.63 \\ %Photometry
 8   & $-$1.85 & +1.11    & +0.38 & +0.04   & \nd   &  \nd    & $-$0.17 & +0.56   & +0.84 \\ %Photometry
11   & $-$1.73 & +0.54    & +0.08 & +0.17   & \nd   &  \nd    & $-$0.10 & +0.35   & +0.77 \\ %Photometry
16   & $-$1.94 & +0.75    & +0.26 & $-$0.26 & \nd   &  \nd    & +0.02   & +0.30   & +0.93 \\ %Photometry
23   & $-$1.69 & $-$0.02  & +0.19 & \nd     & \nd   &  \nd    & $-$0.07 & \nd     & +0.78 \\ %Photomety
31   & $-$1.66 & +0.89    & +0.18 & $-$0.15 & +0.46 &  \nd  & $-$0.18   & +0.54   & +0.77    %Photometry
\enddata
%=================================================================================
\end{deluxetable}

\newpage
\begin{deluxetable}{lll}
\tablewidth{0pt}
\tablecolumns{3}
\tablecaption{Mean Cluster Abundances \label{means}}
\tablehead{
            \colhead{Element Ratio} & \colhead{NGC 6752}  & \colhead{M80}
}
%============
\startdata
\ [Fe/H]             & $-1.58$(0.16)\tablenotemark{a} & $-1.73$(0.13) \\
\ [Al/Fe]            & $+0.51$(0.36) & $+0.37$(0.43) \\
\ [Ca/Fe]            & $+0.26$(0.08) & $+0.24$(0.10) \\
\ [Ti/Fe]$_{\rm I}$  & $+0.16$(0.17) & $+0.12$(0.24) \\
\ [Ti/Fe]$_{\rm II}$ & $+0.40$(0.23) & $+0.46$(0.16) \\
\ [Cr/Fe]            & $-0.05$(0.02) & $-0.02$(0.17) \\
\ [Ni/Fe]            & $-0.12$(0.12) & $-0.05$(0.10) \\
\ [La/Fe]            & $+0.13$(0.14) & $+0.38$(0.15) \\
\ [Eu/Fe]            & $+0.55$(0.12) & $+0.80$(0.12)
\enddata
%============
\tablenotetext{a}{Standard deviation}
\end{deluxetable}

\end{document}